\documentclass[a4paper,10pt,reqno]{amsart}
\DeclareMathSizes{10}{8}{7}{5}
\usepackage[top=1.25in,bottom=1.25in,left=1.20in,right=1.20in]{geometry}
\usepackage{amsmath}
\usepackage {graphicx}
\usepackage{hyperref}
\usepackage[superscript]{cite}
\makeatletter 
\renewcommand\section{\@startsection{section}{1}{\z@}%
{-3.5ex \@plus -1ex \@minus -.2ex}%
{2.3ex \@plus.2ex}%
{\normalfont\normalsize\bf}}
\makeatother

\begin{document}
\title {Correction to Solution of Dirac Equation}
\author {Rui Chen}
\address{Shenzhen Institute of Physics and Mathematics, Shenzhen, 518028, China}
\email{r.ch@163.com}

\begin{abstract}Using the China Unitary Principle to test the Dirac theory
for the hydrogen atomic spectrum shows that the standard Dirac function with
the Dirac energy levels is only one the formal solutions of the
Dirac-Coulomb equation, which conceals some pivotal mathematical
contradictions. The theorem of existence of solution of the Dirac equation
requires an important modification to the Dirac angular momentum constant
that was defined by Dirac's algebra. It derives the modified radial Dirac
equation which has the consistency solution involving the quantum neutron
radius and the neutron binding energy. The inevitable solution for other
atomic energy states is only equivalent to the Bohr solution. It concludes
that the Dirac equation is more suitable to describe the structure of
neutron. How to treat the difference between the unitary energy levels and
the result of the experimental observation of the atomic spectrums for the
hydrogen atom needs to be solved urgently. 
\end{abstract}

\maketitle

\section{Introduction}

The Dirac
equation\cite{Dirac:1928,Dirac:1962,Darwin:1928,Gordon:1928}
was widely used to treat the problems of modern physics. Even in that of
mathematics and chemistry, it also has an outsized effect. However, the
standard Dirac wave function for the hydrogen atom is, in effect, one of the
formal solutions of the Dirac differential equation because of the hidden
mathematical contradictions like the wave function is divergent at the
origin of the coordinate system so that violates the boundary condition
itself. If a wave equation describes precisely the natural phenomena, there
must be the consistency solution\cite{Chen:2008} satisfying the
mathematical rules and physical meaning. In the mathematical perspective of
the Dirac-Coulomb equation, the formal solution comes from the rough
boundary condition without considering the radius of an atomic nucleus,
trying the exact boundary condition with the radius of an atomic nucleus in
place of the traditional rough boundary condition yields the consistency
solution involving the new formula of discrete energy
levels\cite{Chen:2008}. The various solutions of the Dirac equation,
however, still conceal another mathematical difficulty to distinguish the
true from the false. This issue touch on why and how to construct the
constant of angular momentum in the Dirac relativistic quantum mechanics.

Following our work on the Dirac equation in 2008\cite{Chen:2008}, we
here first formulate five mathematical contradictions concealed in the
standard solution of the Dirac equation for the hydrogen atom in detail. As
one of the most important methods testing the mathematical procedures of
theoretical physics we have used from 1985, the new universal principle
called the China Unitary Principle is further introduced to provide great
help for finding the consistency solution of the Dirac equation. We enlarge
the physical meaning of the wave function to the density of the orbit
number. So the quantum mechanics is found to should not be given the meaning
of God play dice\cite{Stewart:2002,Saunders:2000}. Based on the
uniqueness and existence theorem of the differential equations, we determine
the undetermined parameter corresponding to the angular quantum number
instead of the artificial quantum number in the Dirac quantum theory. The
modified form of the radial Dirac equation is a inevitable deduction in
mathematics and physics. It hence gives the inevitable solution of the Dirac
equation that is consistency. We come to a hard mathematical logic for
solving all mathematical difficulties hidden in the standard Dirac theory,
with the result that the exact quantum neutron radius and the neutron
binding energy are obtained\cite{Kim:2006} .

\section{Contradictions hidden in standard solution to Dirac equation}

The new universal principle (why called the China Unitary Principle is to
distinguish from the other unitary principles) is feasible and practical to
reach a consensus on treating those recognized theories concealing the
mathematical difficulties. \textit{Different metrologies can be chosen to describe
the natural laws. Since transforms among different metrologies are certain, and the
laws of nature do not change per se for choosing different metrologies, when different
mathematical forms for describing the same natural laws in different metrologies are
transformed into the same metrologies, they must be the same as the form in the present
metrologies (1=1), which means the transformation is unitary}\cite{Chen:2011}. This principle can help
up to find all logic contradictions hidden in science
theory\cite{Smolin:2007,Woit:2006} as well as mathematics and derive
the consistency solution of the mathematical physical
equation\cite{Chen:2000,Chen:2012} in the end.

It is well known that the Dirac wave equation laid the foundations for
quantum electrodynamics describing the motion and spin of electrons. As a
matter of fact, it was first considered to succeed in describing the fine
structure of the hydrogen atom. We have a lively interest in the strict
solution of the Dirac equation because some new practical theorems of the
optimum differential equations\cite{Chen:2001,Chen:2003} applying to
it gives the completely different result. This is one aspect of violation to
the China Unitary Principle. We looked in particular at the divergence of the
standard Dirac wave function for the $S$ state for the hydrogen atom. In
principle, once a wave equation with the conditions for determining the
solution is given, the only remaining problem for treating the problem of
the quantum mechanics would be a question of pure mathematics. A wave function
concealing a divergent point usually conceals some other mathematical and physical
contradictions. In quantum mechanics, treating the Dirac equation with a Coulomb potential
in method of mathematical physics was finally ascribed to solve a system of radial
differential equations for the upper and lower components of the wave function
\begin{equation}
\label{eq1}
\left[ {c{\rm {\bf \alpha }}\cdot {\rm {\bf \hat {p}}}+\left(
{{\begin{array}{*{20}c}
 {1\;} \hfill & {\;0} \hfill \\
 {0\;} \hfill & {-1} \hfill \\
\end{array} }} \right)mc^2-\frac{Ze^2}{4\pi \varepsilon _0 r}} \right]\left[
{\frac{1}{r}\left( {{\begin{array}{*{20}c}
 F \hfill \\
 G \hfill \\
\end{array} }} \right)} \right]=E\left[ {\frac{1}{r}\left(
{{\begin{array}{*{20}c}
 F \hfill \\
 G \hfill \\
\end{array} }} \right)} \right]
\end{equation}
where ${\rm {\bf \alpha }}$ is the Dirac matrix, ${\rm {\bf \hat
{p}}}=-i\hbar \nabla $, $\hbar =h \mathord{\left/ {\vphantom {h {2\pi }}}
\right. \kern-\nulldelimiterspace} {2\pi }$ with the plank constant $h$, $c$
the velocity of light in a vacuum, and $m$the rest mass of electron. ${\rm
{\bf \alpha }}\cdot {\rm {\bf \tilde {p}}}$ is defined by the Dirac algebra
\begin{equation}
\label{eq2}
{\mathbf{\alpha }} \cdot {\mathbf{\tilde p}} = \left( {\begin{array}{*{20}{c}}
  0&{ - i} \\
  i&0
\end{array}} \right)\left[ { - i\hbar \frac{\partial }{{\partial r}} - \frac{{i\hbar }}{r} + \frac{{i\hbar }}{r}\left( {\begin{array}{*{20}{c}}
  1&0 \\
  0&{ - 1}
\end{array}} \right)\overset{\lower0.5em\hbox{$\smash{\scriptscriptstyle\frown}$}}{\kappa } } \right]
\end{equation}
We are really confused that I have been unable to prove the above expression being an inevitable mathematics deduction.
Where, $\kappa =\pm 1,\;\pm 2,\;\;\cdots \;$ is actually the artificial angular momentum constant. The conventional wisdom states that the equation
(\ref{eq1}) should be given the boundary condition
\begin{equation}
\label{eq3}
\mathop {\lim }\limits_{r\to 0} \frac{1}{r}\left( {{\begin{array}{*{20}c}
 F \hfill \\
 G \hfill \\
\end{array} }} \right)\ne \left( {{\begin{array}{*{20}c}
 {\pm \infty } \hfill \\
 {\pm \infty } \hfill \\
\end{array} }} \right),\quad \frac{1}{r}\left( {{\begin{array}{*{20}c}
 {F\left( {0<r<\infty } \right)} \hfill \\
 {G\left( {0<r<\infty } \right)} \hfill \\
\end{array} }} \right)\ne \left( {{\begin{array}{*{20}c}
 {\pm \infty } \hfill \\
 {\pm \infty } \hfill \\
\end{array} }} \right),\quad \mathop {\lim }\limits_{r\to \infty }
\frac{1}{r}\left( {{\begin{array}{*{20}c}
 F \hfill \\
 G \hfill \\
\end{array} }} \right)=0
\end{equation}
This fixed solution problems has been considered to have the following
unique
solution\cite{Bjorken:1964,Schiff:1968,Thaller:1992,Greiner:2000}
\begin{equation}
\label{eq4}
\frac{1}{r}\left( {{\begin{array}{*{20}c}
 F \hfill \\
 G \hfill \\
\end{array} }} \right)=\left( {{\begin{array}{*{20}c}
 {e^{-\frac{\sqrt {m^2c^4-E_{n_r }^2 } }{\hbar c}r}\sum\limits_{\nu =0}^{n_r
} {b_\nu \left( {\frac{\sqrt {m^2c^4-E_{n_r }^2 } }{\hbar c}r}
\right)^{\sqrt {K^2-Z^2\alpha ^2} +\nu -1}} } \hfill \\
 {e^{-\frac{\sqrt {m^2c^4-E_{n_r }^2 } }{\hbar c}r}\sum\limits_{\nu =0}^{n_r
} {d_\nu \left( {\frac{\sqrt {m^2c^4-E_{n_r }^2 } }{\hbar c}r}
\right)^{\sqrt {K^2-Z^2\alpha ^2} +\nu -1}} } \hfill \\
\end{array} }} \right)
\end{equation}
where the energy is expressed by the symbol $E_{n_r } $ with the subscript
$n_r $ to speak volumes for the mathematical and physical significance,
$\alpha ={e^2} \mathord{\left/ {\vphantom {{e^2} {\left( {4\pi \varepsilon
_0 \hbar c} \right)}}} \right. \kern-\nulldelimiterspace} {\left( {4\pi
\varepsilon _0 \hbar c} \right)}$ is the fine structure constant, and the
undetermined coefficients $b_\nu $ and $d_\nu $ are given by a system of
recurrence relations
\begin{equation}
\label{eq5}
\begin{aligned}
& Z\alpha b_\nu +\left( {\kappa +\sqrt {\kappa ^2-Z^2\alpha ^2} +\nu }
\right)d_\nu -\sqrt {\frac{mc^2-E_{n_r } }{mc^2+E_{n_r } }} b_{\nu -1}
-d_{\nu -1} =0 \\
& \left( {\kappa -\sqrt {\kappa ^2-Z^2\alpha ^2} -\nu } \right)b_\nu +Z\alpha
d_\nu +b_{\nu -1} +\sqrt {\frac{mc^2+E_{n_r } }{mc^2-E_{n_r } }} d_{\nu -1}
=0 \\
 \end{aligned}
\end{equation}
the boundary condition $\mathop {\lim }\limits_{r\to \infty }
\frac{1}{r}\left( {{\begin{array}{*{20}c}
 F \hfill \\
 G \hfill \\
\end{array} }} \right)=0$ requires that the energy $E_{n_r } $ take the
separated value
\begin{equation}
\label{eq6}
E_{\left( {n_r ,K} \right)} =\frac{mc^2}{\sqrt {1+\frac{Z^2\alpha ^2}{\left(
{n_r +\sqrt {\kappa ^2-Z^2\alpha ^2} } \right)^2}} }
\end{equation}
where $n_r =0,1,2,\cdots $ is the radial quantum number.

The equation (\ref{eq4}), (\ref{eq5}) and (\ref{eq6}) are called the standard solution of the
Dirac-Coulomb equation. Because the formula of the energy levels of (\ref{eq6})
agrees with the experimental observation, the standard solution of the Dirac
equation with a Coulomb potential has been considered as succeeding in
describing the fin-structure of the hydrogen-like atoms. However, the
standard solution is not the inevitable deduction of the original Dirac
equation, and it conceals the manifold contradictions\cite{Pestka:2003}:

a) For $Z>137$, as $\kappa =\pm 1$, the energy of the $S$ state given by the
Dirac formula (\ref{eq6}) becomes the imaginary number. This is a well-known
contradiction that has not been solved in the traditional theory. Being a
basic model of mathematical physics, we have not found a legitimate reason
why the Dirac equation should not describe the quantum law of a system
composed by a positive charge with the atomic number larger than 137 and an
electron;

b) For $\kappa =\pm 1$, as $r\to 0$, the standard Dirac wave function (\ref{eq4})
for the $S$ state becomes $\mathop {\lim }\limits_{r\to 0} \frac{1}{r}\left(
{{\begin{array}{*{20}c}
 F \hfill \\
 G \hfill \\
\end{array} }} \right)=\infty $, implying that the standard solution of the
Dirac-Coulomb equation dissatisfies the boundary condition. This leads to a
abnormal statement that all electrons outside the nucleus would rapidly fall
into the atomic nucleus to become the neutron or neutron-like, constructing
the quantum trap that is totally opposed to the structure of matter in the
solar system at least;

c) For $n_r =0$, the formula of the energy eigenvalue (\ref{eq6}) does not hold for
the artificial eigenvalue $\kappa =-1,-2,\cdots $. This can be seen by the
following procedure. Inserting the radial quantum number $n_r =0$ into the
relations (\ref{eq5}) yields $\left( {\kappa -\sqrt {\kappa ^2-Z^2\alpha ^2} }
\right)b_0 +Z\alpha d_0 =0$ and $\left( {mc^2-E_0 } \right)b_0 +\sqrt
{m^2c^4-E_0^2 } d_0 =0$. Eliminating the undetermined coefficients $b_0 $
and $d_0 $ reads a positive equation $\kappa -\sqrt {\kappa ^2-Z^2\alpha ^2}
=Z\alpha \sqrt {{\left( {mc^2-E_0 } \right)} \mathord{\left/ {\vphantom
{{\left( {mc^2-E_0 } \right)} {\left( {mc^2+E_0 } \right)}}} \right.
\kern-\nulldelimiterspace} {\left( {mc^2+E_0 } \right)}} >0$, therefore,
$\kappa >0$. It denotes that the artificial eigenvalue $\kappa =\pm 1,\pm
2,\cdots $ does not hold for all traditional solutions of Dirac equations;

d) The Dirac quantum number $\kappa =\pm \left( {j+1 \mathord{\left/
{\vphantom {1 2}} \right. \kern-\nulldelimiterspace} 2} \right)$ with $j=1
\mathord{\left/ {\vphantom {1 2}} \right. \kern-\nulldelimiterspace} 2,\;3
\mathord{\left/ {\vphantom {3 2}} \right. \kern-\nulldelimiterspace}
2,\;\cdots $ is not a necessarily deduction of the Dirac equation in
mathematics and physics. Defining $j=l\pm 1 \mathord{\left/ {\vphantom {1
2}} \right. \kern-\nulldelimiterspace} 2$ yields immediately the real value
$\kappa =l$ and $\kappa =-\left( {l+1} \right)$, which are now combined into
one expression $\kappa =\pm l$ with $l=0,\;1,\,2,\;3,\cdots $. For $S$
state, because $l=0$, the Dirac formula (\ref{eq6}) becomes an imaginary number.
However, this contradiction has been covered up by a mathematical concept
switch, writing the real deduction into another form $\kappa =\pm \left(
{j+1 \mathord{\left/ {\vphantom {1 2}} \right. \kern-\nulldelimiterspace} 2}
\right)$ with the forced definition $j=1 \mathord{\left/ {\vphantom {1 2}}
\right. \kern-\nulldelimiterspace} 2,\;3 \mathord{\left/ {\vphantom {3 2}}
\right. \kern-\nulldelimiterspace} 2,\;\cdots $, this cannot be strictly
derived in logic actually.

e) The Dirac algebra for constructing the angular momentum constant actually
conflict with theorem of existence of solution for the differential
equations. Because the Dirac equation is relativistic wave equation, the
corresponding angular momentum constant should be first determined by the
eigenvalue of the corresponding relativistic angular operator or the Dirac
equation itself instead of artificial imagination.

These five hidden contradictions show that the standard Dirac wave function
is only one of the formal solutions of the Dirac-Coulomb equation. How to
treat those useful deductions from the paradoxical solution of the wave
equations is an urgent problem.

\section{Statistical interpretation of orbital wave function}

From the first-order Dirac-Coulomb equations (\ref{eq1}) we derived directly the
standard second-order Dirac-Coulomb equations for the hydrogen atom
\begin{equation}
\label{eq7}
\begin{aligned}
& \rho \left( {\alpha -\frac{c_2 }{a}\rho } \right)\frac{d^2G}{d\rho
^2}+\alpha \frac{dG}{d\rho }+\left[ {\frac{\kappa c_2 }{a}+\left( {\alpha
-\frac{c_2 }{a}\rho } \right)^2\left( {\frac{\alpha }{\rho }+\frac{c_1 }{a}}
\right)-\kappa ^2\left( {\frac{\alpha }{\rho }-\frac{c_2 }{a}} \right)}
\right]G=0 \\
& \rho \left( {\alpha +\frac{c_1 }{a}\rho } \right)\frac{d^2F}{d\rho
^2}+\alpha \frac{dF}{d\rho }+\left[ {\frac{\kappa c_1 }{a}+\left( {\alpha
+\frac{c_1 }{a}\rho } \right)^2\left( {\frac{\alpha }{\rho }-\frac{c_2 }{a}}
\right)-\kappa ^2\left( {\frac{\alpha }{\rho }+\frac{c_1 }{a}} \right)}
\right]F=0 \\
 \end{aligned}
\end{equation}
where $c_1 ={\left( {mc^2+E} \right)} \mathord{\left/ {\vphantom {{\left(
{mc^2+E} \right)} {\hbar c}}} \right. \kern-\nulldelimiterspace} {\hbar
c},\;c_2 ={\left( {mc^2-E} \right)} \mathord{\left/ {\vphantom {{\left(
{mc^2-E} \right)} {\hbar c}}} \right. \kern-\nulldelimiterspace} {\hbar
c},\;a=\sqrt {c_1 c_2 } ,\;\rho =ar$. One can only obtain the formal formula
of energy eigenvalue for the hydrogen-like atoms by solving the above stand
second-order Dirac equation without considering the existence and uniqueness
of solutions, but the formal wave function of the second-order differential
equations (\ref{eq7}) is not equivalent to the formal wave function of the original
first-order differential equations (\ref{eq1}). One the other hand, the second-order
differential equations (\ref{eq7}) are not the optimum differential equations,
implying that the formal solutions do not satisfy the differential equations
actually. In this logic connection, the first-order Dirac-Coulomb equation
is chosen as one metrologies and the second-order Dirac-Coulomb equation is
chosen as another metrologies. It is incontestably apparent that the
solutions of the standard second-order Dirac-Coulomb equation (\ref{eq7}) and the
solution of the first-order Dirac-Coulomb equation (\ref{eq1}) destroy the Chen
unitary principle. Actually, it has been proven that all kinds of
second-order Dirac equations and all kinds of imitated first-order Dirac
equations have no consistency solutions satisfying the boundary condition
(\ref{eq3}). To tackle these problems, one has to consider again the boundary
condition of the wave equations, which is often misunderstood in physics.

Born's statistical
interpretation\cite{Born:1955,Pais:1982,Ballentine:1970} on
the wave function has been universally accepted. It is found that the
probability postulation is fully compatible with the classical orbit.
Considering an artificial satellite turning around the earth, for example,
its probability density can be translated as the density of the orbit number
(the orbit number per unit volume). In fact, classical mechanics theory and
quantum mechanics theory are just two metrologies for describing the order
of nature. The concept of the classical orbit will not suddenly disappear
because of quantum mechanics theory. According to the Chen unitary
principle, using the probability amplitudes for the position of the particle
to determine the orbit of the particle is necessarily equivalent to using
the density of the orbit number to determine the probability of finding the
particle. FIG. 1 shows symbolically the responding relation between the wave
function's probability distribution and the motion orbit of an electron
outside the nucleus of an atom. For the same energy level, an electron may
have the different circle or the ellipse orbits. Its orbit plane is varying
continuously because of the electromagnetic disturbance. The farther the
electron is away from the nuclear, the smaller the probability of the
electron crossing a given surface appearing to be. The electrons absorb and
radiate the energy to produce the orbit transition. At the surface of an
atomic nuclear, the probability of an electron crossing must be the certain
limited value but not $\infty $. From this macroscopic picture of
statistical interpretation on the wave function, it is not difficult to find
that the quantum mechanics should not give the meaning of God play dice.

\begin{figure}[htbp]
\centerline{\includegraphics[width=2.95in,height=2.95in]{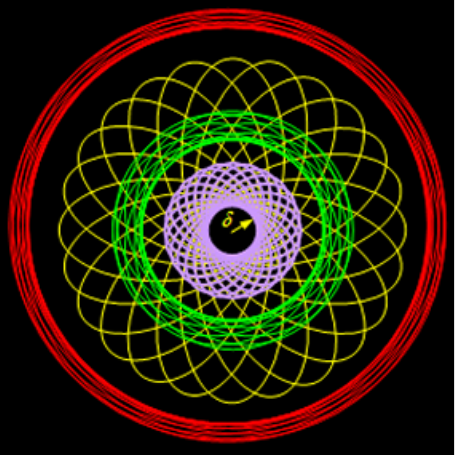}}
\label{fig1}
\end{figure}

\begin{quote}
\small FIG.1. Schematic drawing for the electron's
probability density and motion orbits. Each group of ellipse curves
of different colours that express the orbit of the electron
corresponds to the certain special energy states. The density of
curves with the speed weight of moving electron is proportional to
the magnitude of the probability density of the electron.
\end{quote}
\normalsize

The initial-boundary value conditions play an important role for determining
the consistency solution of a wave equation from its general solutions. In
order to overcome the divergence of relativistic wave functions at the
origin\cite{Malenfant:1987}, it should consider the quantum radius of the
atomic nucleus. As one of the reliable treatments, we assume that equation
(\ref{eq4}) holds only for  $r\ge \delta $. Inside
$\delta $, the potential has to be modified from a Coulomb ${Ze^2}
\mathord{\left/ {\vphantom {{Ze^2} {4\pi \varepsilon _0 \delta }}} \right.
\kern-\nulldelimiterspace} {4\pi \varepsilon _0 \delta }$ potential to one
corresponding to a spread-out charge distribution. Therefore, to do a
completely correct calculation, one solves the Dirac equation separately
outside of and inside of $\delta $, with two different potentials, and then
matches, at $r=\delta $, the outside solution (i.e., the standard
Dirac-Coulomb one) to the inside solution. The inside solution is the finite constant;
its effect is to modify the energy-level formula by a small
correction that takes into account the finite radius of the nuclear. This
idea just supports introducing the exact boundary condition to the wave
equations for hydrogen-like atom.
\begin{equation}
\label{eq8}
\mathop {\lim }\limits_{r\to \delta } \frac{1}{r}\left(
{{\begin{array}{*{20}c}
 F \hfill \\
 G \hfill \\
\end{array} }} \right)\ne \left( {{\begin{array}{*{20}c}
 {\pm \infty } \hfill \\
 {\pm \infty } \hfill \\
\end{array} }} \right),\; \frac{1}{r}\left( {{\begin{array}{*{20}c}
 {F\left( {\delta <r<\infty } \right)} \hfill \\
 {G\left( {\delta <r<\infty } \right)} \hfill \\
\end{array} }} \right)\ne \left( {{\begin{array}{*{20}c}
 {\pm \infty } \hfill \\
 {\pm \infty } \hfill \\
\end{array} }} \right),\; \mathop {\lim }\limits_{r\to \infty }
\frac{1}{r}\left( {{\begin{array}{*{20}c}
 F \hfill \\
 G \hfill \\
\end{array} }} \right)=0
\end{equation}
where $\delta $ is the quantum radius of the atomic nucleus.

Further, investigating the existence of solution of the Dirac equation leads
to an important modification to the Dirac theory, and the quantum neutron
radius and the neutron binding energy are hence derived. According to the
China Unitary Principle, if the standard solution of the Dirac equation with
the traditional boundary condition (\ref{eq3}) were correct, then the corresponding
new exact solution of the Dirac equation with the exact boundary condition
(\ref{eq8}) would deduce $\delta =0$. Otherwise, we can only face up to a logically
inevitable consequence of the Dirac equation.

\section{Natural quantum number and consistency solution to Dirac equation}

Any one mathematical contradiction hidden in the standard solution of the
Dirac equation suffice to ignore the standard solution. We note that the
standard Dirac theory lacks the enough logic of both mathematics and physics
to prove that the artificial eigenvalue $\kappa =\pm 1,\,\pm 2,\;\pm
3,\;\cdots $ is inevitable. The inevitable deduction is $\kappa =\pm l\left(
{l=0,\;1,\;2,\;3\cdots } \right)$ which conceals a null result by the
imaginary energy solution for $l=0$. In order to solve the contradiction, we
introduce the inevitable parameter $\mathbb{C}$ to replace the artificial
parameter$_{ }\kappa $, $\mathbb{C}$ is determined by the existence and
uniqueness of solution of the Dirac differential equation with the exact
boundary condition (\ref{eq8}). Hence, the equations (\ref{eq2}) becomes $\vec {\alpha
}\cdot \hat {\vec {p}}=\alpha _r \left[ {-i\hbar \left( {\partial
\mathord{\left/ {\vphantom {\partial {\partial r}}} \right.
\kern-\nulldelimiterspace} {\partial r}+1 \mathord{\left/ {\vphantom {1 r}}
\right. \kern-\nulldelimiterspace} r} \right)+{i\hbar \beta \mathbb{C}}
\mathord{\left/ {\vphantom {{i\hbar \beta \mathbb{C}} r}} \right.
\kern-\nulldelimiterspace} r} \right]$. It could be seen that the modified
form to the radial Dirac equation (\ref{eq1}) is
\begin{equation}
\label{eq9}
\begin{aligned}
\left[ {i\hbar c\left( {\frac{\partial }{{\partial r}} + \frac{1}{r}} \right)\left( {\begin{array}{*{20}{c}}
{0\;}&{ - i}\\
{i\;}&{\;0}
\end{array}} \right) + \frac{{\hbar c}\mathbb{C}}{r}\left( {\begin{array}{*{20}{c}}
{0\;}&1\\
{1\;}&0
\end{array}} \right) - m{c^2}\left( {\begin{array}{*{20}{c}}
{1\;}&{\;0}\\
{0\;}&{ - 1}
\end{array}} \right) + \left( {E + \frac{{Z{e^2}}}{{4\pi {\varepsilon _0}r}}} \right)} \right] \cdot \left[ {\frac{1}{r}\left( {\begin{array}{*{20}{c}}
F\\
G
\end{array}} \right)} \right] = 0
 \end{aligned}
\end{equation}
Let $\xi =r-\delta $, this equation equivalents to a system of differential
equations
\begin{equation}
\label{eq10}
\begin{aligned}
& \left( {\frac{Z\alpha }{\xi +\delta }-\frac{mc^2-E}{\hbar c}}
\right)F+\left( {\frac{\mathbb{C}+1}{\xi +\delta }+\frac{d}{d\xi }}
\right)G=0 \\
& \left( {\frac{Z\alpha }{\xi +\delta }+\frac{mc^2+E}{\hbar c}}
\right)G+\left( {\frac{\mathbb{C}-1}{\xi +\delta }-\frac{d}{d\xi }}
\right)F=0 \\
 \end{aligned}
\end{equation}
On the other hand, an atomic nucleus has the certain quantum radius
corresponding to the exact boundary condition. A correct wave equation
describing the hydrogen-like atom must have the exact solution that
satisfies the exact boundary condition, and the radius of the atomic nucleus
should be obtained. We have derived the general solution of the system of
first-order differential equations (\ref{eq10}) with the exact boundary condition
(\ref{eq8}), it takes the form
\begin{equation}
\label{eq11}
R\left( r \right)=\frac{1}{r}\left( {{\begin{array}{*{20}c}
 F \hfill \\
 G \hfill \\
\end{array} }} \right)=\frac{1}{\xi +\delta }\left( {{\begin{array}{*{20}c}
 {\exp (-\frac{\sqrt {m^2c^4-E_{n_r }^2 } }{\hbar c}\xi )\sum\limits_{\nu
=0}^{n_r } {b_\nu } \xi ^\nu } \hfill \\
 {\exp (-\frac{\sqrt {m^2c^4-E_{n_r }^2 } }{\hbar c}\xi )\sum\limits_{\nu
=0}^{n_r } {d_\nu } \xi ^\nu } \hfill \\
\end{array} }} \right)
\end{equation}
where the coefficients $b_\nu $ and $d_\nu $ satisfy the following system of
recurrence relations\cite{Chen:2008}
\begin{equation}
\label{eq12}
\begin{array}{l}
 \frac{E_{n_r } -mc^2}{\hbar c}b_{\nu -1} +\left( {\frac{E_{n_r }
-mc^2}{\hbar c}\delta +\alpha } \right)b_\nu -\frac{\sqrt {m^2c^4-E_{n_r }^2
} }{\hbar c}d_{\nu -1} +\left( {\mathbb{C}+\nu -\frac{\sqrt {m^2c^4-E_{n_r
}^2 } }{\hbar c}\delta } \right)d_\nu =0 \\
 \frac{\sqrt {m^2c^4-E_{n_r }^2 } }{\hbar c}b_{\nu -1} +\left(
{\mathbb{C}-\nu +\frac{\sqrt {m^2c^4-E_{n_r }^2 } }{\hbar c}\delta }
\right)b_\nu +\frac{E_{n_r } +mc^2}{\hbar c}d_{\nu -1} +\left( {\frac{E_{n_r
} +mc^2}{\hbar c}\delta +\alpha } \right)d_\nu =0 \\
 \end{array}
\end{equation}
the eigenvalue of the energy for hydrogen-like atom is easily obtained to be
\begin{equation}
\label{eq13}
E_{n_r } =\frac{mc^2}{\sqrt {1+\frac{Z^2\alpha ^2}{n_r^2 }} }\mbox{ }\left(
{n_r =0,1,2,3,\cdots } \right)
\end{equation}
Different from the formal energy levels derived via the radial Dirac
equation with the rough boundary condition and the artificial angular
momentum constant (\ref{eq6}), the inevitable energy level determined by the modified
radial Dirac equation with the exact boundary condition. $E_{n_r } $ is
independent of the intrinsic angular quantum number $\mathbb{C}$. This
result denotes that the Dirac equation cannot describe the fine-structure of
the hydrogen-like atoms.

The above wave function (\ref{eq10}) with (\ref{eq12}) and (\ref{eq13}) are the consistency solution
of the Dirac equation for the hydrogen-like atom without the divergence
point of wave function. Introducing the intrinsic angular quantum number
$\mathbb{C}$ to the radial Dirac equation is critical. Otherwise, the
existence of solution to the Dirac equation would be destroyed. In the
inevitable solution, there is not any contradiction concealed in the standard
solution. Because of various mathematical contradictions, the traditionally
formal solution of the Dirac equation with the rough boundary condition,
should be given up.

Quantum mechanics shall never bring the quantum trap that all electrons
outside of the nucleus would fall into the atomic nucleus because the wave
function divergence. Otherwise all matter in the universe would exist as
neutron, but it was not true. Consequently, a wave function should have no
patience with any divergence. Now, we discuss the special solution of
Dirac-Coulomb equation for $S$ state. Inserting $n_r =0$ into the recurrence
relations (\ref{eq12}) reads
\begin{equation}
\label{eq14}
\begin{aligned}
& \left( {\frac{mc^2-E_0 }{\hbar c}\delta -Z\alpha } \right)b_0 -\left(
{\mathbb{C}_0 -\frac{\sqrt {m^2c^4-E_0^2 } }{\hbar c}\delta } \right)d_0 =0
\\
& \left( {\mathbb{C}_0 +\frac{\sqrt {m^2c^4-E_0^2 } }{\hbar c}\delta }
\right)b_0 +\left( {\frac{mc^2+E_0 }{\hbar c}\delta +Z\alpha } \right)d_0 =0
\\
& \frac{mc^2-E_0 }{\hbar c}b_0 +\frac{\sqrt {m^2c^4-E_0^2 } }{\hbar c}d_0 =0
\\
& \frac{\sqrt {m^2c^4-E_0^2 } }{\hbar c}b_0 +\frac{mc^2+E_0 }{\hbar c}d_0 =0
\\
 \end{aligned}
\end{equation}
In this system of recurrence relations, the last two relations are
equivalent. By the first and second relations, we obtain the quantum radius
of hydrogen-like atomic nucleus $\delta ={\hbar c\left( {\mathbb{C}_0^2
-Z^2\alpha ^2} \right)} \mathord{\left/ {\vphantom {{\hbar c\left(
{\mathbb{C}_0^2 -Z^2\alpha ^2} \right)} {\left( {2Z\alpha E_0 } \right)}}}
\right. \kern-\nulldelimiterspace} {\left( {2Z\alpha E_0 } \right)}$.
Substituting the first relation into the third or fourth relation reads $E_0
={\left( {\mathbb{C}_0^2 -Z^2\alpha ^2} \right)mc^2} \mathord{\left/
{\vphantom {{\left( {\mathbb{C}_0^2 -Z^2\alpha ^2} \right)mc^2} {\left(
{\mathbb{C}_0^2 +Z^2\alpha ^2} \right)}}} \right. \kern-\nulldelimiterspace}
{\left( {\mathbb{C}_0^2 +Z^2\alpha ^2} \right)}$. This formula of energy
eigenvalue belongs to the atomic nucleus. Combining the above two formulas
gives $\delta ={\left( {\mathbb{C}_0^2 +Z^2\alpha ^2} \right)\hbar }
\mathord{\left/ {\vphantom {{\left( {\mathbb{C}_0^2 +Z^2\alpha ^2}
\right)\hbar } {\left( {2Z\alpha mc} \right)}}} \right.
\kern-\nulldelimiterspace} {\left( {2Z\alpha mc} \right)}$. In order to give
the exact value of radius of hydrogen-like atomic nucleus and their energy
eigenvalue for special state corresponding to $n_r =0$, it must needs
determine the parameter $\mathbb{C}_0 $. Substituting the second formula
into the third formula in (\ref{eq14}) gives $\mathbb{C}_0 =Z\alpha \sqrt {{\left(
{mc^2-E_0 } \right)} \mathord{\left/ {\vphantom {{\left( {mc^2-E_0 }
\right)} {\left( {mc^2+E_0 } \right)}}} \right. \kern-\nulldelimiterspace}
{\left( {mc^2+E_0 } \right)}} $. Inserting it into $E_0 ={\left(
{\mathbb{C}_0^2 -Z^2\alpha ^2} \right)mc^2} \mathord{\left/ {\vphantom
{{\left( {\mathbb{C}_0^2 -Z^2\alpha ^2} \right)mc^2} {\left( {\mathbb{C}_0^2
+Z^2\alpha ^2} \right)}}} \right. \kern-\nulldelimiterspace} {\left(
{\mathbb{C}_0^2 +Z^2\alpha ^2} \right)}$ gives $E_0 =0$. One obtains the
formal solution $\mathbb{C}_0 =\pm Z\alpha $, but $\mathbb{C}_0 =-Z\alpha $
dissatisfies the relations (\ref{eq14}) and should be abnegated. The physical
solution is $\mathbb{C}_0 =Z\alpha $. Substituting it into the expression
$\delta ={\left( {\mathbb{C}_0^2 +Z^2\alpha ^2} \right)\hbar }
\mathord{\left/ {\vphantom {{\left( {\mathbb{C}_0^2 +Z^2\alpha ^2}
\right)\hbar } {\left( {2Z\alpha mc} \right)}}} \right.
\kern-\nulldelimiterspace} {\left( {2Z\alpha mc} \right)}$ yields $\delta
={Z\alpha \hbar } \mathord{\left/ {\vphantom {{Z\alpha \hbar } {mc}}}
\right. \kern-\nulldelimiterspace} {mc}$, namely
\begin{equation}
\label{eq15}
\delta =\frac{Ze^2}{4\pi \varepsilon _0 mc^2}
\end{equation}
this is the quantum neutron-like radius. The special eigenvalue $E_0 =0$
denotes the unique energy state of the neutron-like. For $Z=1$, the formula
(\ref{eq13}) reads the neutron binding energy
\begin{equation}
\label{eq16}
\Delta E=E_\infty -E_0 =mc^2
\end{equation}
where $E_\infty $ is the energy of the hydrogen atom corresponding to $n_r
=\infty $ in the formula (\ref{eq13}), and $E_0$ is the energy of a neutron..

In physical and mathematical ideas, (\ref{eq11}), (\ref{eq12}), (\ref{eq13}), (\ref{eq16}), (\ref{eq15}) satisfy the
China Unitary Principle. They form the real solution of the radial Dirac
equation with a Coulomb potential.

The neutron binding energy implies that the neutron can be broken up by a
photon of the energy $m_e c^2$, or perhaps an electron and a proton could
combine into a neutron and emit a photon of the energy $m_e c^2$ at the same
time. Usually, $\delta ={e^2} \mathord{\left/ {\vphantom {{e^2} {4\pi
\varepsilon _0 mc^2}}} \right. \kern-\nulldelimiterspace} {4\pi \varepsilon
_0 mc^2}$ (=2.8117940285 fm) is regarded as the classical electron radius.
The above analysis shows that it should be the quantum neutron radius. This
is about triplication of the neutron radius, which the recent value is
reported to be 0.8418 fm, the early results are 0.805(\ref{eq12})\cite{Hand:1963} ,
0.861(20) fm\cite{Bourzeix:1996} , 0.862(\ref{eq13}) fm\cite{Simon:1980} , 0.8768
fm, 0.88014 fm\cite{Rosenfelder:2000} , 0.89014 fm and 0.895$\pm
$0.018~fm\cite{Sick:2003} and so on. These data are actually calculated by
Lamb shift\cite{Lamb:1950,Weitz:1994,Pachucki:1994} .

\section{Summary and expectation}

In the present paper, we enlarge the physical meaning of the wave function
to the density of the orbit number, implying that the quantum mechanics
should not give the meaning of God play dice, and Schrodinger Cat may do not
constitute a real problem of science. We showed that the Dirac equation with
a Coulomb potential has two different exact solutions. One is the formal
solution consists of the Dirac formula of the energy eigenvalue which
actually describes the structure of the hydrogen atom, and another is the
consistency solution consists of the both neutron binding energy and atom
energy level which actually describes the Bohr atom structure as well as the
neutron. It suggested not to break any of the rules of mathematics and
physics to solve the relativistic wave equation.

The formal solution of the Dirac equation results from the rough boundary.
It conceals five contradictions destroying the fundamental mathematical
rules but has been covered up by some abstract exploitations or mathematical
technique. For example, the divergence of Dirac wave function at the origin
for $S$ state was called the ``mild divergence''. Because the formal
solution derives the famous relativistic correction of the energy eigenvalue
for the hydrogen atom, all exploitations fencing with the mathematical
contradiction have been accepted. On the statistical interpretation to the
wave function in view of macroscopic orbit concept, replacing the rough
boundary condition by the exact boundary condition and replacing the
artificial angular momentum eigenvalue $\kappa =\pm 1,\pm 2,\cdots $ by the
inevitable eigenvalue $\mathbb{C}$ are the mainly amelioration for the Dirac
theory. The result shows that the consistency solution of the Dirac equation
can foreshow the neutron binding energy and the quantum neutron radius. We
conclude that the Dirac equation is more suitable to describe the structure of neutron.

No infallibility will be claimed for the correction to the solution for the
Dirac equations. It may need further improvement in quantum mechanics. The
consistency solution of the Dirac equation involves the inevitable formula
energy level that is worth only the Bohr energy level. Why cannot the
consistency solution of the Dirac equation with the exact boundary condition
describe the fine structure of the hydrogen atom? We do not think those
formal theories that can give the solutions of expectancy are the ultimate
theories of physics. The question as well as the faster than light
discoveries\cite{Wang:2000,Adeam:2011,Geoff:2011} raise new
possibilities for treating the Dirac equation a second
time\cite{Schiller:2001} even an exact amelioration to the wave
equation\cite{Blair:2008}\cite{Blair:2009}\cite{Metcalfe:2005} , in which
the mathematical rigour will not be destroyed.

\section*{Acknowledgments}
We express my thanks to Dr. Jerome Malenfant (Senior Assistant Editor PRL)
for one of his emails about the exact boundary condition of the hydrogen atom.
We gratefully acknowledge Professor Dr. Michael H. Brill (Principal Color Scientist of U.S.A.)
for his profound discussions and some important contributions to the application of the China Unitary Principle.
We would like to acknowledge the help in language of Caitlin Jackson, Emily Hsieh,
Robert Montalbano (U.S.A.) and Jiewen Liang (Shenzhen Foreign Languages School of China).

\end{document}